\documentstyle[aps,prl,multicol]{revtex}
\input psfig

\begin{document}

\title{Slow dynamics of water under pressure}

\author{Francis~W. Starr$^{1}$, Stephen Harrington$^{1}$, Francesco
Sciortino$^{2}$, and H. Eugene Stanley$^{1}$}

\address{$^1$Center for Polymer Studies, Center for Computational
Science, and Department of Physics, \\ Boston University, Boston, MA
02215 USA}

\address{$^2$Dipartmento di Fisica e Instituto Nazionale per la Fisica
della Materia, \\ Universit\'{a} di Roma ``La Sapienza'', Piazzale Aldo
Moro 2, I-00185, Roma, Italy}

\date{Submitted: December 31, 1998}
 
\maketitle

\begin{abstract}
We perform lengthy molecular dynamics simulations of the SPC/E model of
water to investigate the dynamics under pressure at many temperatures
and compare with experimental measurements.  We calculate the isochrones
of the diffusion constant $D$ and observe power-law behavior of $D$ on
lowering temperature with an apparent singularity at a temperature
$T_c(P)$, as observed for water.  Additional calculations show that the
dynamics of the SPC/E model are consistent with slowing down due to the
transient caging of molecules, as described by the mode-coupling theory
(MCT).  This supports the hypothesis that the apparent divergences of
dynamic quantities along $T_c(P)$ in water may be associated with
``slowing down'' as described by MCT.
\end{abstract}

\begin{multicols}{2}

On supercooling water at atmospheric pressure, many thermodynamic and
dynamic quantities show power-law growth~\cite{debenedetti}.  This power
law behavior also appears under pressure, which allows measurement of
the locus of apparent power-law singularities in water
[Fig.~\ref{fig:water-T_s}(a)].  The possible explanations of this
behavior have generated a great deal of interest.  In particular, three
scenarios have been considered: (i) the existence of a spinodal bounding
the stability of the liquid in the superheated, stretched, and
supercooled states \cite{spinodal}; (ii) the existence of a
liquid-liquid transition line between two liquid phases differing in
density \cite{pses,critical-point,ms98}; (iii) a singularity-free
scenario in which the thermodynamic anomalies are related to the
presence of low-density and low-entropy structural heterogeneities
\cite{singfree}.  Based on both
experiments~\cite{prielmeier,germans,xenon} and recent
simulations~\cite{francesco}, several authors have suggested that the
power-law behavior of dynamic quantities might be explained by the
transient caging of molecules by neighboring molecules, as described by
the mode-coupling theory (MCT)~\cite{mct}, which we address here.  This
explanation would indicate that the dynamics of water are explainable in
the same framework developed for other fragile liquids~\cite{angell95},
at least for temperatures above the homogeneous nucleation temperature
$T_H$.  Moreover, this explanation of the dynamic behavior on
supercooling may be independent of the above scenarios suggested for
thermodynamic behavior [Fig.~\ref{fig:water-T_s}(a)].

Here we focus on the behavior of the diffusion constant $D$ under
pressure, which has been studied experimentally~\cite{prielmeier}.  We
perform molecular dynamics simulations in the temperature range 210~K --
350~K for densities ranging from 0.95~g/cm$^3$ -- 1.40~g/cm$^3$
[Table~\ref{table:state-points}] using the extended simple point charge
potential (SPC/E)~\cite{spce}.  We select the SPC/E potential because it
has been previously shown to display power-law behavior of dynamic
quantities, as observed in supercooled water at ambient
pressure~\cite{francesco,bc94}.

In Fig.~\ref{fig:D(P)}, we compare the behavior of $D$ under pressure at
\hfill several \hfill temperatures \hfill for \hfill our \hfill
simulations \hfill and \hfill the

\newbox\figa
\setbox\figa=\psfig{figure=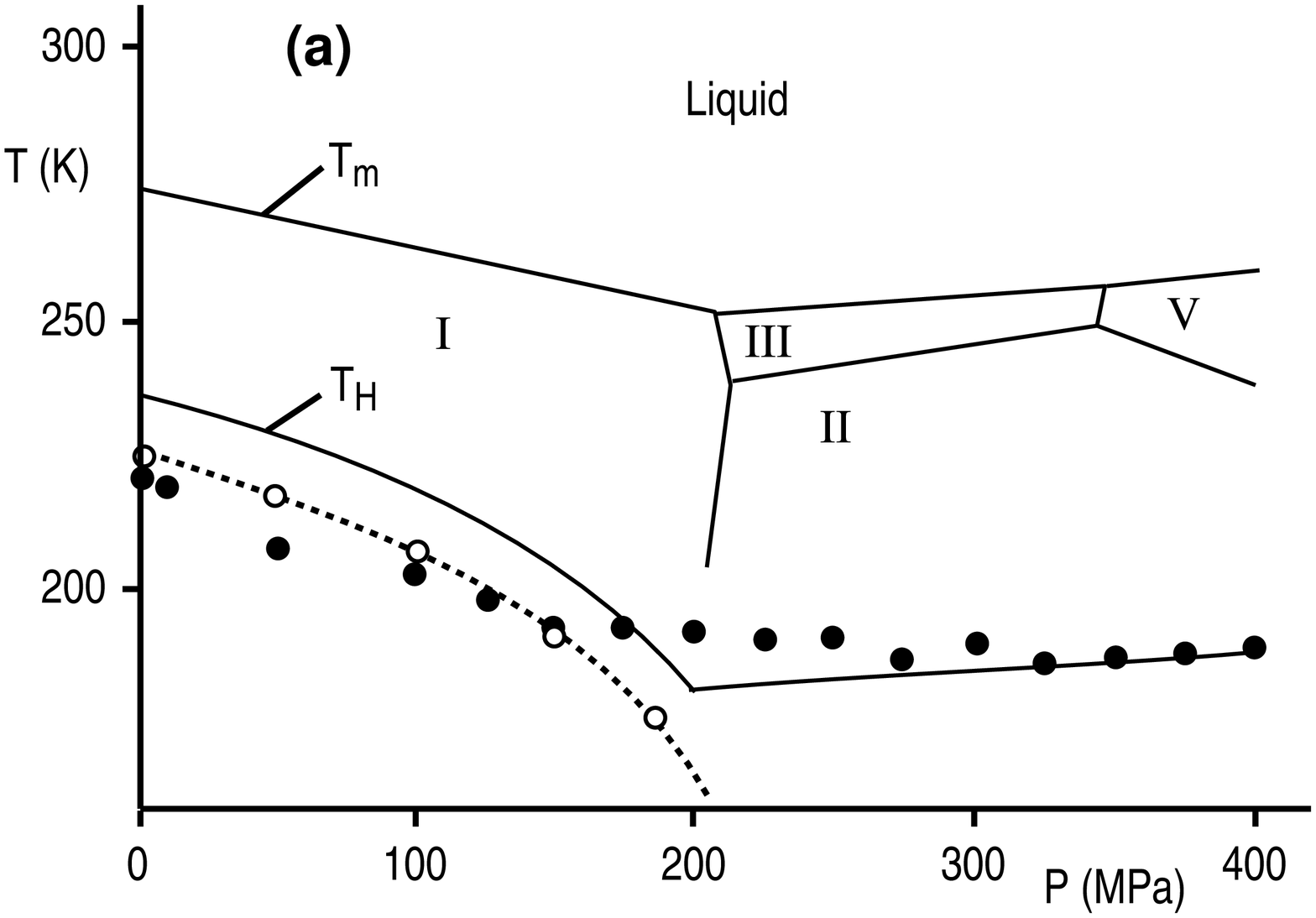,width=3.05in} \newbox\figb
\setbox\figb=\psfig{figure=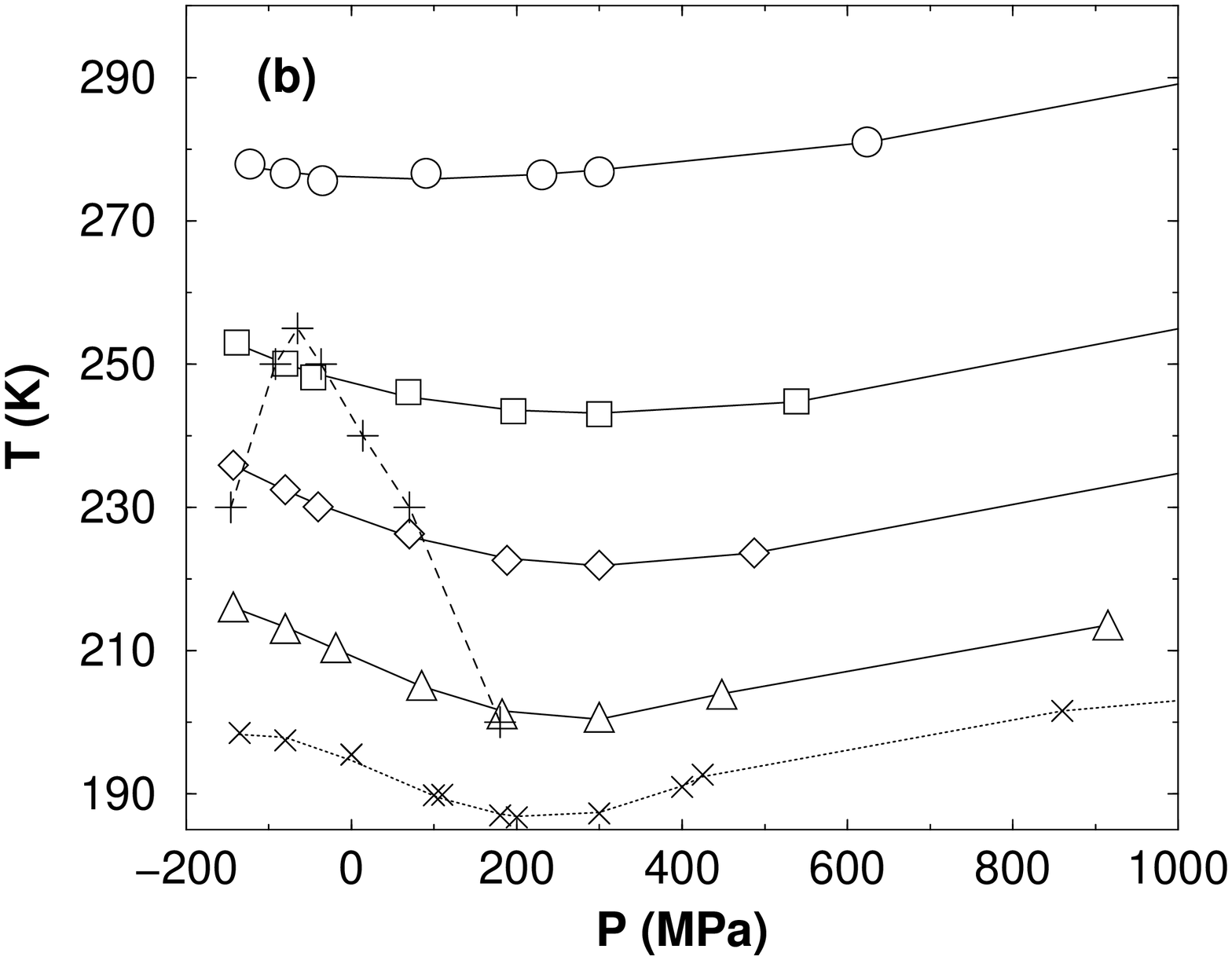,width=3.35in,angle=-90}
\begin{figure*}
\begin{center}
\leavevmode
\centerline{\box\figa}
\centerline{\box\figb}
\narrowtext
\caption{(a) Phase diagram of water.  The extrapolated
divergence of the isothermal compressibility
($\circ$)~\protect\cite{kanno-angell79} and the extrapolated divergence
of $D$ (filled $\circ$)~\protect\cite{prielmeier}.  The different
locations of these divergences suggest that the phenomena may arise from
different explanations.  (b) Isochrones of $D$ from simulation.  The
lines may be identified at follows: $D = 10^{-5}$cm$^2$/s ($\circ$); $D
= 10^{-5.5}$~cm$^2$/s ($\Box$); $D = 10^{-6}$cm$^2$/s ($\diamond$); $D =
10^{-7}$cm$^2$/s ($\triangle$).  The diffusion is also fit to $D \sim
(T/T_c-1)^\gamma$.  The locus of $T_c$ is indicated by ($\times$).  For
reference, the ($+$) symbols indicate the locus of T$_{\mbox{\scriptsize
MD}}$ found in ref.~\protect\cite{hpss97}.  }
\label{fig:water-T_s}
\end{center}
\end{figure*}

\newbox\figa
\setbox\figa=\psfig{figure=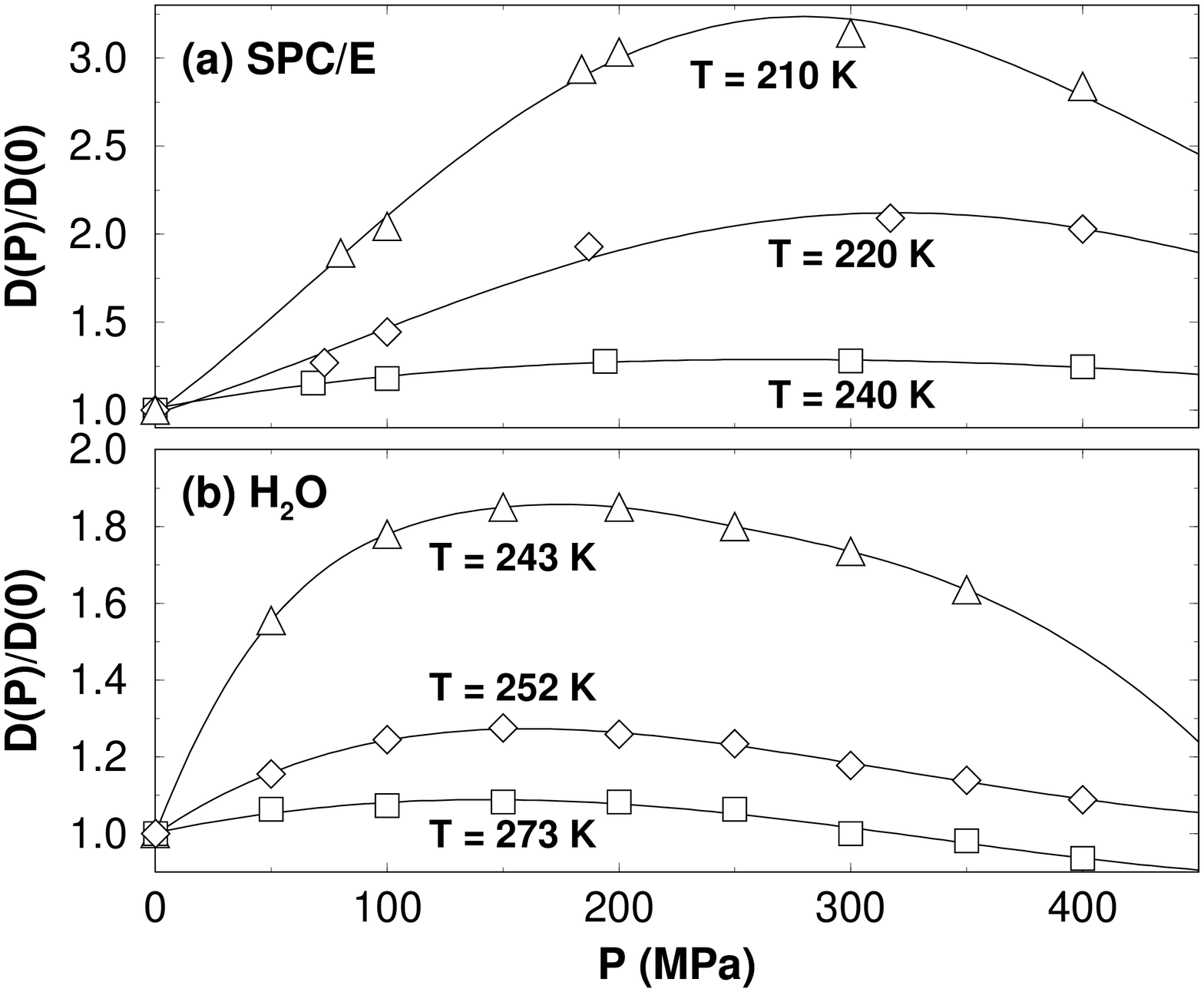,width=3.35in,angle=-90}
\begin{figure*}[htbp]
\begin{center}
\leavevmode
\centerline{\box\figa}
\narrowtext
\caption{Diffusion constant $D$ as a function of pressure for various
temperatures from (a) our simulations and (b) NMR studies of
water~\protect\cite{prielmeier}.  }
\label{fig:D(P)}
\end{center}
\end{figure*}

\noindent
experiments of ref.~\cite{prielmeier}.  The anomalous increase in $D$ is
qualitatively reproduced by SPC/E, but the quantitative increase of $D$
is significantly larger than that observed experimentally.  This
discrepancy may arise form the fact that the SPC/E potential is {\it
under-structured} relative to water~\cite{hpss97}, so applying pressure
allows for more bond breaking and thus greater diffusivity than observed
experimentally.  We also find that the pressure where $D$ begins to
decrease with pressure -- normal behavior for a liquid -- is larger than
that observed experimentally.  This simple comparison of $D$ leads us to
expect that the qualitative dynamic features we observe in the SPC/E
potential will aid in the understanding of the dynamics of water under
pressure, but will likely not be quantitatively accurate.

We next determine the approximate form of the lines of constant $D$
(isochrones) by interpolating our data over the region of the phase
diagram studied [Fig.~\ref{fig:water-T_s}(b)].  We note that the locus
of points where the slope of the isochrones changes sign (i.e. the locus
of points where $D$ obtains a maximum value) is close to the
$T_{\mbox{\scriptsize MD}}$ locus~\cite{hpss97}.  At each density
studied, we fit $D$ to a power law $D \sim (T/T_c - 1)^{\gamma}$.  The
shape of the locus of $T_c$ values compares well with that observed
experimentally~\cite{prielmeier}, and changes slope at the same pressure
[Figs.~\ref{fig:water-T_s}(a) and (b)].  We find the striking feature
that $\gamma$ decreases under pressure for the SPC/E model, while
$\gamma$ increases experimentally [Fig.~\ref{fig:gamma}].  This
disagreement underscores the need to improve the dynamic properties of
water models, most of which already provide an adequate account of
static properties~\cite{ha98}.

We next consider interpretation of our results using MCT, which has been
used to quantitatively describe the weak supercooling regime -- i.e.,
the temperature range \hfill where \hfill the \hfill characteristic
\hfill times \hfill become \hfill three \hfill or 

\newbox\figa
\setbox\figa=\psfig{figure=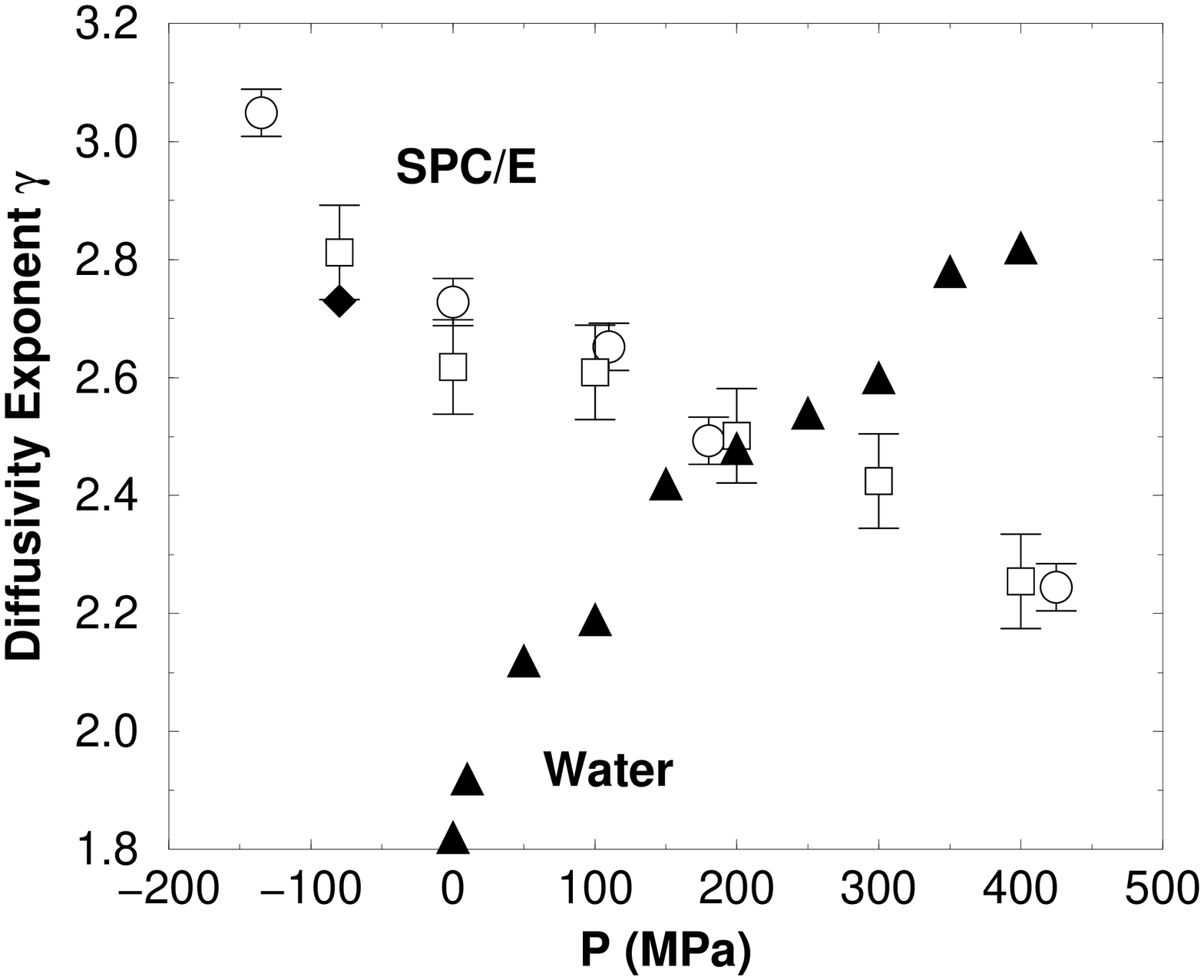,width=3.35in,angle=-90}
\begin{figure*}[htbp]
\begin{center}
\leavevmode
\centerline{\box\figa}
\narrowtext
\caption{Pressure dependence of the diffusivity exponent $\gamma$
defined by $D \sim (T/T_c-1)^\gamma$.  The symbols may be identified as
follows: ($\circ$) $\gamma$ measured from simulation along isochores;
($\Box$) $\gamma$ measured from simulation along isobaric paths, which
are estimated from the isochoric data; ($\Diamond$) $\gamma$ measured
along the -80~MPa isobar in ref.~\protect\cite{francesco}; ($\triangle$)
experimental measurements of $\gamma$ in water from
ref.~\protect\cite{prielmeier}.  It is clear from the available data
that the SPC/E potential fails to reproduce the qualitative behavior of
$\gamma$ under pressure in liquid water.}
\label{fig:gamma}
\end{center}
\end{figure*}

\noindent four orders of magnitude larger than those of the normal
liquid~\cite{ediger}.  The region where experimental data are available
in supercooled water is exactly the region where MCT holds.  MCT
provides a theoretical framework in which the slowing down of the
dynamics arises from caging effects, related to the coupling between
density modes, mainly over length scales on the order of the nearest
neighbors.  In this respect, MCT does not require the presence of a
thermodynamic instability to explain the power-law behavior of the
characteristic times.

MCT predicts power-law behavior of $D$, and also that the Fourier
transform of the density-density correlation function $F(q,t)$,
typically referred to as the intermediate scattering function, decays
via a two-step process.  $F(q,t)$ can be measured by neutron scattering
experiments and is calculated via

\begin{equation}
F(q,t) \equiv \frac{1}{S(q)} \left\langle \sum_{j,k=1}^N e^{-i {\bf
q}\cdot[{\bf r}_k(t) - {\bf r}_j(0)]} \right\rangle ,
\label{eq:isf}
\end{equation}

\noindent
where $S(q)$ is the structure factor~\cite{hansen-mcdonald}.  In the
first relaxation step, $F(q,t)$ approaches a plateau value
$F_{\mbox{\scriptsize plateau}}(q)$; the decay from the plateau has the
form $F_{\mbox{\scriptsize plateau}}(q) - F(q,t) \sim t^b$, where $b$ is
known as the von Schweidler exponent.  According to MCT, the value $b$
is completely determined by the value of $\gamma$~\cite{goetze-lh}, so
calculation of these exponents for SPC/E determines if MCT is consistent
with our results.

\newbox\figa
\setbox\figa=\psfig{figure=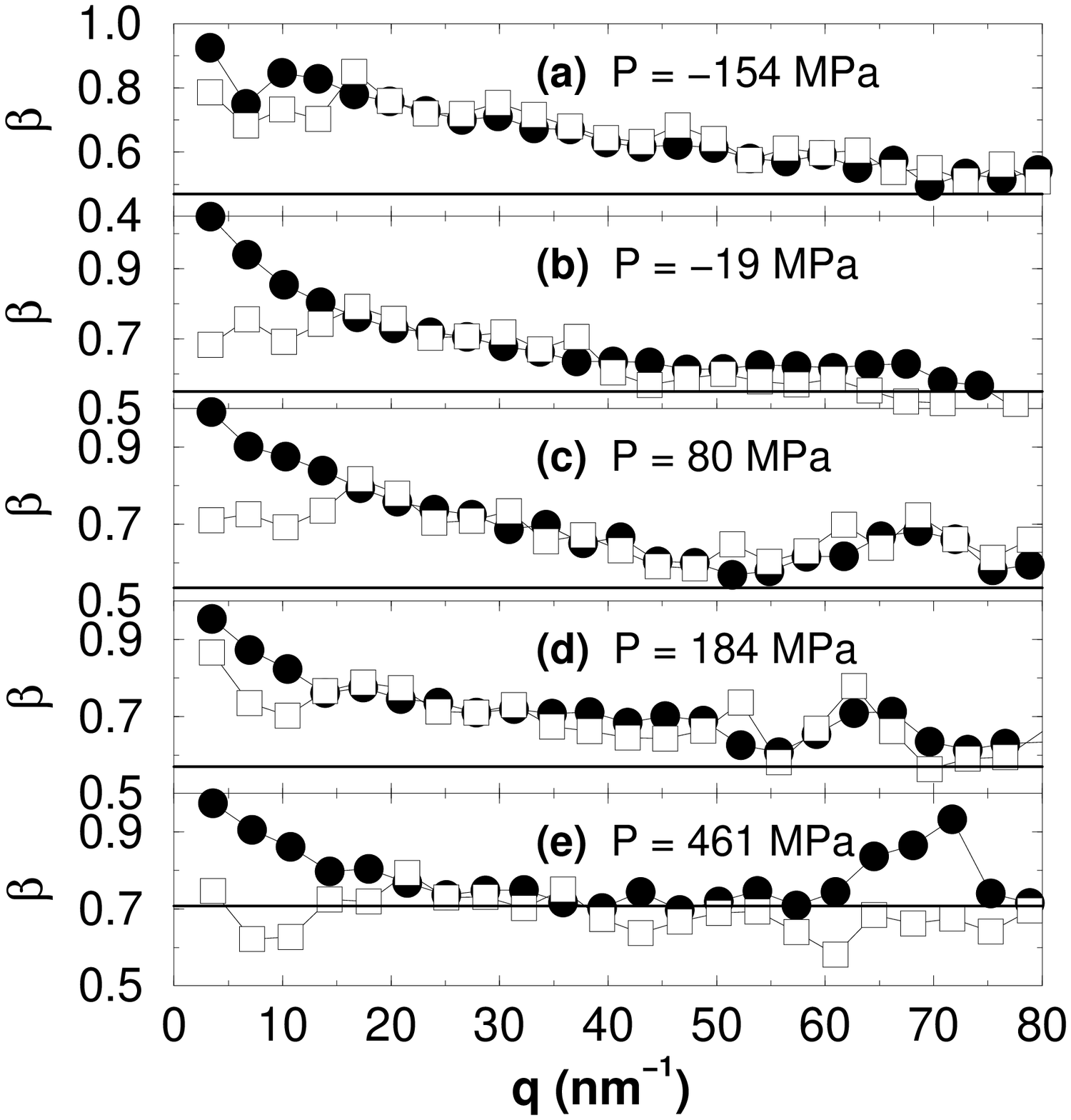,width=3.35in,angle=-90}
\begin{figure*}[htbp]
\begin{center}
\leavevmode
\centerline{\box\figa}
\narrowtext
\caption{Fit of the stretched exponential of Eq.~(\ref{eq:kww}) for $t >
2$~ps at $T=210$~K to both $F_{\mbox{\scriptsize self}}(q,t)$ ($\circ$)
and $F(q,t)$ ($\Box$) to obtain $\beta$.  The horizontal line indicates
the value predicted by MCT for $b$ using $\gamma$ values extrapolated
from Fig.~\ref{fig:gamma}.  For $P \protect\gtrsim 80$~MPa, the
relaxation of $F(q,t)$ for $q \protect\gtrsim 60$~nm$^{-1}$ comes almost
entirely from the first decay region, so the $\beta$ values obtained are
not reliable in this range.  Longer simulations, currently underway,
will produce more reliable results in this region.}
\label{fig:betaq}
\end{center}
\end{figure*}

The range of validity of the power-law $t^b$ is strongly
$q$-dependent~\cite{mayr}, making unambiguous calculation of $b$
difficult.  Fortunately, the same exponent $b$ controls the long-time
behavior of $F(q,t)$ at large $q$. Indeed, MCT predicts that at long
time, $F(q,t)$ decays according to a Kohlrausch-Williams-Watts stretched
exponential

\begin{equation}
F(q,t) = A(q)
\exp{\left[\left(\frac{t}{\tau(q)}\right)^{\beta(q)}\right]},
\label{eq:kww}
\end{equation}

\noindent
with $\lim_{q\to\infty} \beta(q) = b$~\cite{fuchs-beta}.  We show the
$q$-dependence of $\beta$ for each density studied at $T = 210$~K
[Fig.~\ref{fig:betaq}].  We also calculate $\beta$ for the ``self-part''
of $F(q,t)$, denoted $F_{\mbox{\scriptsize self}}(q,t)$~\cite{isf-note}.
In addition, we show the expected value of $b$ according to MCT, using
the values of $\gamma$ extrapolated from Fig.~\ref{fig:gamma}.  The
large-$q$ limit of $\beta$ appears to approach the value predicted by
MCT~\cite{confirmation}.  Hence we conclude that the dynamic behavior of
the SPC/E potential in the pressure range we study is consistent with
slowing down as described by MCT [Fig.~\ref{fig:b-gamma}].  We also note
that on increasing pressure, the values of the exponents become closer
to those for hard-sphere ($\gamma=2.58$ and $b=0.545$) and Lennard-Jones
($\gamma=2.37$ and $b=0.617$) systems~\cite{hs-lj}.  \hfill This \hfill
confirms \hfill that \hfill the \hfill hydrogen-bond \hfill network 

\newbox\figa 
\setbox\figa=\psfig{figure=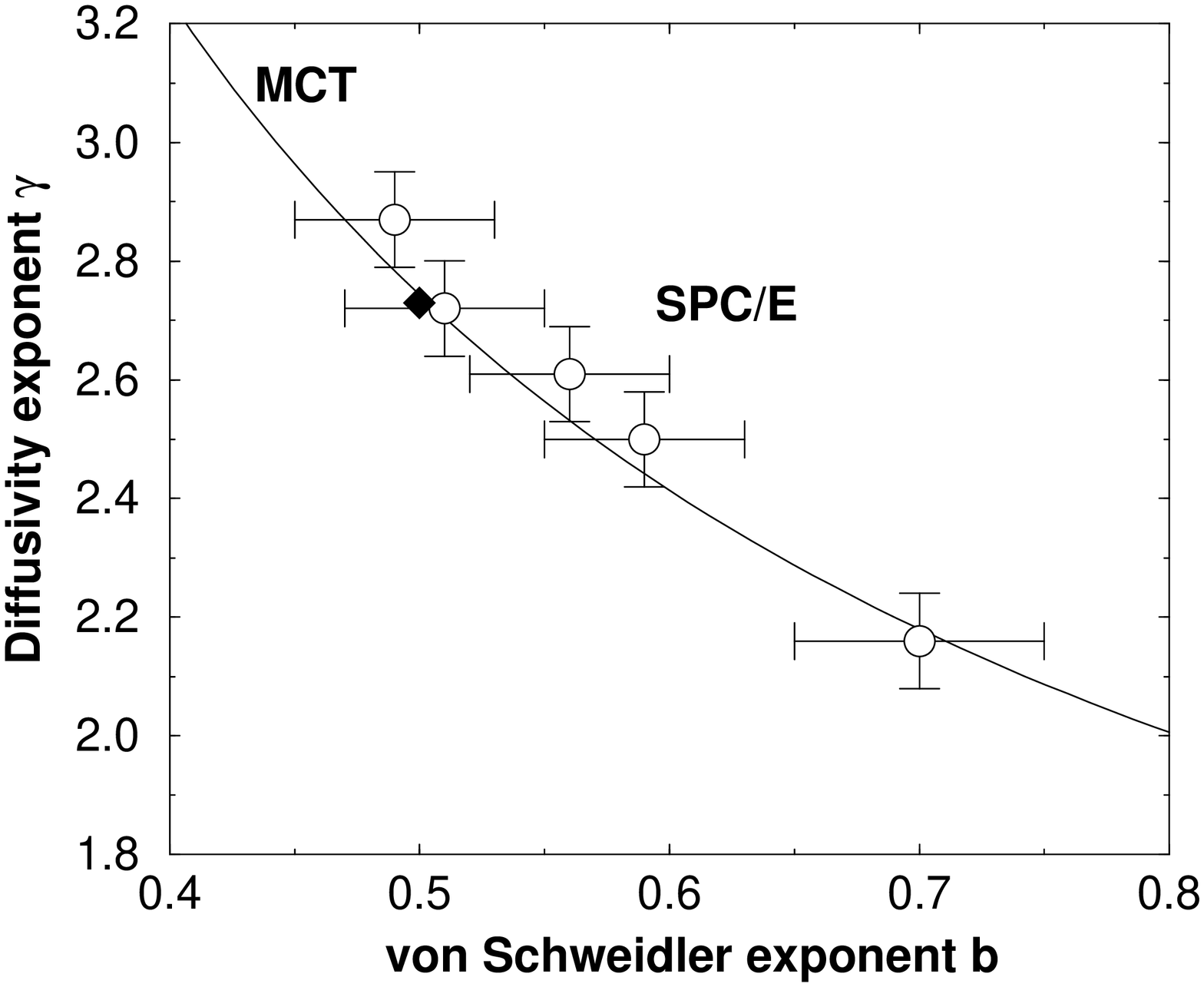,width=3.35in,angle=-90}
\begin{figure*}[htbp]
\begin{center}
\leavevmode
\centerline{\box\figa}
\narrowtext
\caption{The line shows the predicted relationship between $b$ and
$\gamma$ from MCT. The symbols show the calculated values for the SPC/E
model: ($\circ$) from this work, (filled $\Diamond$) from
ref.~\protect\cite{francesco}.}
\label{fig:b-gamma}
\end{center}
\end{figure*}

\noindent is destroyed under pressure and that the water dynamics become
closer to that of normal liquids, where core repulsion is the dominant
mechanism.

A significant result of our analysis is the demonstration that MCT is
able to rationalize the dynamic behavior of the SPC/E model of water at
all pressures.  In doing so, MCT encompasses both the behavior at low
pressures, where the mobility is essentially controlled by the presence
of strong energetic cages of hydrogen bonds, and at high pressures,
where the dynamics are dominated by excluded volume effects.

We wish to thank A.~Rinaldi, S.~Sastry, and A.~Scala for their
assistance.  FWS is supported by an NSF graduate fellowship.  The Center
for Polymer Studies is supported by NSF grant CH9728854 and British
Petroleum.

\end{multicols}

\begin{minipage}{7in}
\begin{center}
\begin{table}
\caption{Summary of the state points simulated.  We simulate 216 water
molecules interacting via the SPC/E pair potential~\protect\cite{spce}.
We simulate two independent systems at all temperatures (except 350~K),
as the large correlation time makes time averaging more difficult.  We
equilibrate all simulated state points to a constant temperature by
monitoring the pressure and internal energy.  We control the temperature
using the Berendsen method of rescaling the
velocities~\protect\cite{ber-scaling} with a thermostat time of 200~ps.
The reaction-field technique with a cutoff of 0.79 nm accounts for the
long-range Coulomb interactions~\protect\cite{rxn-field}.  The equations
of motion evolve using the SHAKE algorithm~\protect\cite{shake} with a
time step of 1 fs.  Additional details can be found in
ref.~\protect\cite{francesco}.  Systems are equilibrated for a time
$t_{\mbox{\scriptsize eq}}$, followed by data collection runs for a time
$t_{\mbox{\scriptsize data}}$.  For all state points, the uncertainty in
the potential energy $U$ is less the 0.05~kJ/mol.}
\medskip
\begin{tabular}{ccccccc}
T & $\rho$ (g/cm$^3$) & U (kJ/mol) & P (MPa) & D ($10^{-6}$ cm$^2$/s) &
$t_{\mbox{\scriptsize eq}}$ (ns) & $t_{\mbox{\scriptsize data}}$ (ns) \\
\tableline
210 & 0.95 & $-53.84 $ & $-154 \pm 9$ & 0.0272 & 25 & 50\\
    & 1.00 & $-53.70 $ & $-19 \pm  11$ & 0.0913 & 35 & 50\\
    & 1.05 & $-53.43 $ & $80  \pm  12$ & 0.214 & 30 & 50\\
    & 1.10 & $-53.24 $ & $184 \pm  13$ & 0.331 & 30 & 50\\
    & 1.20 & $-53.13 $ & $461  \pm 14$ & 0.290 & 25 & 50\\
\tableline
220 & 0.95 & $-53.00$  & $-150 \pm 6$ & 0.168 & 15 & 15 \\
    & 1.00 & $-52.87$  & $-21 \pm 10$ & 0.389 & 15 & 15 \\
    & 1.05 & $-52.73$  & $73 \pm 8$  & 0.558 & 15 & 15 \\
    & 1.10 & $-52.59$  & $187 \pm 8$ & 0.847 & 15 & 15\\
    & 1.20 & $-52.48$  & $480 \pm 9$ & 0.801 & 15 & 15\\
    & 1.30 & $-52.49$  & $951 \pm 12$ & 0.263 & 15 & 15\\
\tableline
240 & 0.95  & $-51.33 $ & $-153  \pm 8$ & 1.41 & 7 & 5\\
    & 1.00  & $-51.35 $ & $-45 \pm   9$ & 1.87 & 7 & 5\\
    & 1.05  & $-51.34 $ & $68 \pm    9$ & 2.44 & 7 & 5\\
    & 1.10  & $-51.28 $ & $195 \pm  10$ & 2.70 & 7 & 5\\
    & 1.20  & $-51.24 $ & $527 \pm  11$ & 2.37 & 7 & 5\\
    & 1.30  & $-51.25 $ & $1035 \pm  4$ & 1.35 & 7 & 5\\
\tableline
260 & 0.95 & $-49.68 $ & $-148 \pm  9$ & 5.04 & 5 & 3\\
    & 1.00 & $-49.87 $ & $-43  \pm 10$ & 6.08 & 5 & 3\\
    & 1.05 & $-49.93 $ & $77   \pm 11$ & 5.91 & 5 & 3\\
    & 1.10 & $-50.00 $ & $212  \pm 11$ & 5.88 & 5 & 3\\
    & 1.20 & $-50.10 $ & $572  \pm 13$ & 5.74 & 5 & 3\\
    & 1.30 & $-50.14 $ & $1127 \pm 14$ & 3.54 & 5 & 3\\
\tableline
300 & 0.95 & $-46.80 $ & $-109 \pm 12$ & 19.9 & 0.5 & 1\\
    & 1.00 & $-47.20 $ & $-13 \pm 13 $ & 20.0 & 0.5 & 1\\
    & 1.05 & $-47.49 $ & $112 \pm 14 $ & 18.3 & 0.5 & 1\\
    & 1.10 & $-47.65 $ & $264 \pm 14 $ & 18.2 & 0.5 & 1\\
    & 1.20 & $-47.95 $ & $678 \pm 16 $ & 15.3 & 0.5 & 1\\
    & 1.30 & $-48.06 $ & $1293 \pm 18$ & 11.2 & 0.5 & 1\\
\tableline
350 & 1.00 & $-44.35 $ & $62  \pm  18$ & 49.7 & 0.5 & 40 ps\\
    & 1.10 & $-45.15 $ & $358 \pm  20$ & 38.1 & 0.5 & 40 ps\\
    & 1.20 & $-45.56 $ & $828 \pm  22$ & 27.0 & 0.5 & 40 ps\\
    & 1.30 & $-45.76 $ & $1504 \pm 25$ & 18.0 & 0.5 & 40 ps\\
\end{tabular}
\label{table:state-points}
\end{table}
\end{center}
\end{minipage}


\begin{references}
\bibitem{isf-note} $F(q,t)$ may be split into two contributions: the
correlations of a molecule with itself [$F_{\mbox{\scriptsize
self}}(q,t)$], and the correlations between pairs of molecules
[$F_{\mbox{\scriptsize distinct}}(q,t)$].  We calculate the
$q$-dependence of $\beta$ for $F_{\mbox{\scriptsize self}}(q,t)$ because
we have much better statistics for the self-correlations than for
cross-correlations.
\bibitem{confirmation} We confirm that the values of $b$ calculated from
$\lim_{q\to\infty} \beta(q) = b$ are consistent with the von Schweidler
power-law.
\end{references}
\end{document}